\documentclass[aps,prl,twocolumn,superscriptaddress,preprintnumbers,nofootinbib,amsmath,amssymb,tighten]{revtex4} 

\usepackage{epsfig}
\usepackage{bm}
\usepackage{color}
\definecolor{Green}{rgb}{0.0, 0.56, 0.20}

\def\beq{\begin{equation}}
\def\eeq{\end{equation}}
\def\bea{\begin{eqnarray}}
\def\eea{\end{eqnarray}}
\def\ben{\begin{enumerate}}
\def\een{\end{enumerate}}
\def\bq{\begin{quote}}
\def\eq{\end{quote}}

\def \gsim{\mathrel{\vcenter
     {\hbox{$>$}\nointerlineskip\hbox{$\sim$}}}}
\def\gappeq{\mathrel{\rlap {\raise.5ex\hbox{$>$}}
{\lower.5ex\hbox{$\sim$}}}}
\def\lappeq{\mathrel{\rlap{\raise.5ex\hbox{$<$}}
{\lower.5ex\hbox{$\sim$}}}}

\def\mec{\mu \! \to \! e~ {\rm conversion}}
\def\meg{\mu \to e \gamma}
\def\teg{\tau \to e \gamma}
\def\tmg{\tau \to \mu \gamma}

\def\a{\alpha}
\def\b{\beta}
\def\g{\gamma}

\def\m{\mu}

\begin{document}

\preprint{LA-UR-17-21718} 
\preprint{OUHEP-17-1}

\title{Spin-dependent $\bm{\mu \to e}$ conversion}

\author{Vincenzo Cirigliano}
\affiliation{Theoretical Division, Los Alamos National Laboratory,
Los Alamos, NM 87545, USA}
\author{Sacha  Davidson}
\affiliation{
 IPNL, CNRS/IN2P3, Universit\'e Lyon 1,Univ. Lyon, 
 69622 Villeurbanne, France  \\ 
 }
\author{Yoshitaka Kuno}
\affiliation{Department of Physics, Osaka University, 
1-1 Machikaneyama, Toyonaka, Osaka 560-0043, Japan}
\date{\today}

\begin{abstract}
The experimental sensitivity to $\mu \to e$ conversion on nuclei is expected 
 to improve by four orders of magnitude in coming years. We consider the impact of  $\mu \to e$ flavour-changing tensor and axial-vector four-fermion operators 
 which couple to the  spin of nucleons. Such operators, which have not previously been considered,  contribute to $\mu\to e$ conversion in three ways: 
in nuclei with spin they mediate a spin-dependent transition;  in all nuclei they contribute to the 
 coherent ($A^2$-enhanced) spin-independent conversion 
via finite recoil effects 
and via  loop mixing with dipole, scalar, and vector operators.   
We estimate the spin-dependent rate in  Aluminium  (the target of the upcoming COMET and Mu2e experiments), show that the loop effects give the greatest sensitivity to  tensor and axial-vector operators involving first-generation quarks, and  discuss the complementarity of the spin-dependent and independent 
contributions to  $\mu \to e$ conversion. 
\end{abstract}

\pacs{Valid PACS appear here}
\maketitle

{\bf Introduction} -- 
New particles   and interactions  beyond the Standard Model
 of particle physics  are required to explain
 neutrino masses and mixing angles.
The   search for traces  of this New Physics (NP)  is
 pursued on many fronts. One possibility
 is to look  directly for the new particles
 implicated in neutrino mass generation,
 for instance 
 at the LHC~\cite{LHC} { or } SHiP~\cite{SHiP}. A  complementary approach
 seeks new interactions among  known
 particles,  such as neutrinoless double
 beta decay~\cite{0nu2B} or Charged Lepton Flavour Violation (CLFV)~\cite{KO}.

CLFV transitions  of charged leptons  
are induced by the observed  massive neutrinos,
at  unobservable rates  suppressed by $(m_\nu/m_W)^4 \sim 10^{-48}$.
 A detectable rate  would  point to the existence of 
  new heavy particles, as may arise in
 models that generate neutrino  masses, or
  that address other puzzles of the Standard Model
 such as the hierarchy problem. Observations of CLFV
 are therefore  crucial  to identifying the
 NP of the lepton sector, providing information
 complementary to direct searches. 

{
From a theoretical perspective, 
at energy scales well  below the masses
 of  the new particles, CLFV can be
 parametrised with effective operators
  (see  {\it e.g.}~\cite{Georgi}),
 constructed out of the kinematically accessible
 Standard Model (SM) fields, and respecting
 the  relevant gauge symmetries. 
 In this effective field theory (EFT) description,  
 information about the underlying new dynamics 
 is encoded in the operator coefficients, calculable 
 in  any given model.
}
 
The experimental sensitivity to 
 a wide variety of CLFV processes 
is systematically improving. Current bounds on 
branching ratios of 
 $\tau$  flavour changing decays such as 
 $\tmg$, $\teg$  and $\tau \to 3 \ell$~\cite{Babartmgteg,Belletmg,Belletmmm}
 are ${\cal O} (10^{-8})$, and Belle-II is expected
 to improve the sensitivity by an order of magnitude~\cite{Belle-2}. The bounds
 on  the  $\mu \leftrightarrow e$ flavour changing processes
 are currently of order  $\sim 10^{-12}$~\cite{Bertl:2006up,Bellgardt:1987du}, 
 with the most restrictive contraint from the
 MEG collaboration: $BR(\meg) \leq 4.2 \times 10^{-13}$~\cite{TheMEG:2016wtm}.
 { Future experimental
   sensitivities should improve by several orders of magnitude,
   in particular, 
 the COMET~\cite{COMET} and  Mu2e~\cite{mu2e} 
  experiments   
 aim to  reach a   sensitivity
 to $\mec$  on nuclei  of $\sim 10^{-16}$,
  and  the PRISM/PRIME proposal\cite{PP} could
 reach the unprecedented level of $10^{-18}$.}

In searches for $\mec$, a  $\mu^-$
 from the beam is captured by a nucleus in the target, and
tumbles down to the $1s$ state. The muon
will  be  closer to the nucleus than an electron ($ r \sim \alpha Z/m$),
due to its larger mass. In the presence of  a CLFV interaction
with the quarks that compose the nucleus,
or with its electric field, the muon can transform
into an electron. This electron,
emitted with an energy $E_e \simeq  m_\mu $, is the signature of
$\mec$.

Initial analytic estimates of  the $\mec$ rate were obtained
by  Feinberg and Weinberg~\cite{FeinWein},
a wider range of  nuclei were studied numerically by
Shankar~\cite{Shanker:1979ap}, and relativistic
effects relevant in heavier nuclei were included
in Ref.~\cite{czarM2}.   State of the art conversion rates 
for  a broad range of nuclei
induced by CLFV operators
which  can  contribute coherently to $\mec$  were obtained in  Ref.~\cite{KKO}, while 
some missing  operators were included in Ref.~\cite{CKOT}.

The calculation has some similarities with dark
matter scattering on nuclei~\cite{JKG,BBPS,EPV}, where the  cross-section 
can be  classified as
spin-dependent (SD) or spin-independent (SI). 
Previous analyses of $\mec$~\cite{KKO,CKOT} focused on CLFV
interactions involving a scalar or
vector nucleon current, because,
similarly to SI dark matter scattering,
these sum coherently across the nucleus
at the amplitude level, giving 
an amplification $\sim A^2$ in the rate,
where $A$ is the atomic number.
However, other  processes are possible, 
such as  spin-dependent conversion on the
ground state nucleus, which we explore here, or  incoherent $\mec$, where the 
final-state nucleus is in an excited state~\cite{Kosmas:1993ch,Shanker:1979ap}.

The  upcoming exceptional  experimental sensitivities motivate
our study of new  contributions to $\mec$ 
induced by tensor and axial vector operators\footnote{
 We  leave out the light-quark pseudoscalar operators 
and gluon operators such as $G \tilde{G}$ that can be induced 
by  heavy-quark pseudoscalar operators at the heavy quark thresholds. 
The effect of this class of  operators  in a nucleus is 
suppressed both by spin  and momentum transfer.}, 
 which were  not considered in  Refs.~\cite{KKO,CKOT}.  
These operators couple to the  spin of the nucleus and  
can induce ``spin-dependent'' $\mec$ in nuclei with spin (such as Aluminium, the
proposed target of COMET and Mu2e),  not enhanced by $A^2$. 
In addition,  the tensor and axial operators will
contribute  to   ``spin-independent''    
conversion via finite-momentum-transfer corrections~\cite{FHKLX,etalCirelli}, 
 and Renormalisation Group  mixing~\cite{megmW,PSI}~\footnote{The analogous mixing of SD to SI  dark matter interactions
was discussed in~\cite{Uli,Crivellin:2014qxa}.}.
{ In  an EFT  framework,  
our analysis shows new sensitivities   to  
previously
unconstrained combinations of dimension-six operator coefficients,
as we illustrate below.
In the absence of CLFV,
this gives new constraints
on the coefficients,
and when
CLFV is observed, it
could assist in
determining its origin.}

{\bf  Estimating the $\mu \to e$ conversion rate} --  
Our starting point is the effective Lagrangian~\cite{KO}  
\bea
\delta {\cal L} & = & - 2\sqrt{2} G_F
\sum_{Y} {\Big (} C_{D,Y} {\cal O}_{D,Y}
+ C_{GG,Y} {\cal O}_{GG,Y} 
\nonumber \\
&& 
+ \sum_{q= u,d,s} 
\sum_{O} C_{O,Y}^{qq} {\cal O}_{O,Y}^{qq} 
+h.c. {\Big )}
\label{LVAPST}
\eea
where  $Y \in\{L,R\}$ and  $O \in\{V,A,S,T\}$ 
and the operators are explicitly given by ($P_{L,R} = 1/2 (I \mp \gamma_5)$)
\bea
    {\cal O}_{D,Y}& = &m_\mu (\overline{e} \sigma^{\a \b} P_Y \mu )F_{\a \b}
     \nonumber \\ {\cal O}_{GG,Y}& =& \frac{9}{32 \pi^2 m_t} (\overline{e}  P_Y \mu ){\rm Tr}[G_{\a \b} G^{\a\b}] \nonumber \\
{\cal O}^{ qq}_{V,Y} &=& (\overline{e} \gamma^\a P_Y \mu )  (\overline{q} \gamma_\a  q )  \nonumber \\
{\cal O}^{ qq}_{A,Y} &= &(\overline{e} \gamma^\a P_Y \mu )  (\overline{q} \gamma_\a \g_5 q )  \nonumber \\
{\cal O}^{ qq}_{S,Y} &=& (\overline{e}  P_Y \mu )  (\overline{q}  q ) 
\nonumber \\
{\cal O}^{ qq}_{T,Y} &=&  (\overline{e} \sigma^{\a \b} P_Y \mu )  (\overline{q} \sigma_{\a \b}  q  )~ .
\label{opsdefn}
\eea
While our primary focus is on the  tensor 
(${\cal O}^{ qq}_{T,Y}$)  
and axial  (${\cal O}^{ qq}_{A,Y}$) operators, 
we include the vector, scalar, dipole and gluon  operators 
because  the first three are induced by loops, and
the last arises  by integrating out heavy quarks.

At zero momentum transfer, the quark bilinears can be matched onto  nucleon  bilinears 
\bea
 \bar{q}(x) \Gamma_O q(x) \rightarrow G^{N,q}_O  \bar{N}(x) \Gamma_O N(x)
\label{G}
\eea
where the vector charges are $G^{p,u}_{V} = G^{n,d}_{V} = 2$ and 
$G^{p,d}_{V} = G^{n,u}_{V} = 1$, and for the  axial charges 
we use the results inferred in Ref.~\cite{BBPS} 
by using the HERMES measurements~\cite{HERMES}, namely
$G^{p,u}_A = G^{n,d}_A=  0.84(1)$, 
$G^{p,d}_A = G^{n,u}_A = -0.43(1)$,  and 
$G^{p,s}_A = G^{n,s}_A = -.085(18)$. 
For the tensor charges we use the lattice QCD results~\cite{ClatticePRL} 
in the $\overline{\rm MS}$ scheme  at $\mu = 2$~GeV, namely  
$G^{p,u}_T = G^{n,d}_T=  0.77(7)$, $G^{p,d}_T = G^{n,u}_T = -0.23(3)$, 
and   $G^{p,s}_T = G^{n,s}_T = .008(9)$,
Finally, { for the scalar charges induced by light quarks 
we use  a precise  dispersive determination~\cite{Hoferichter:2015dsa},} 
$G^{p,u}_S = \frac{m_N}{m_u} 0.021(2)$, $G^{p,d}_S =  \frac{m_N}{m_d}  0.041(3) $, 
$G^{n,u}_S = \frac{m_N}{m_u} 0.019(2) $, and  $G^{n,d}_S = \frac{m_N}{m_d} 0.045(3) $, 
and  an average of lattice results~\cite{Junnarkar:2013ac}
for the strange charge:
$G^{p,s}_S =    G^{n,s}_S =    \frac{m_N}{m_s}  0.043(11) $.
In  all cases,  
 we take central values of the 
 $\overline{\rm MS}$  quark masses  at $\mu = 2$~GeV, 
 namely $m_u = 2.2$~MeV,   $m_d = 4.7$~MeV, and  $m_s = 96$~MeV~\cite{Agashe:2014kda}.

Taking the above matching into account, the nucleon-level effective Lagrangian has the 
same structure of  (\ref{LVAPST})  with the replacements $\bar{q} \Gamma_O q \to \bar{N} \Gamma_O N$ 
and with effective couplings given by~\footnote{
  The gluon operators ${\cal O}_{GG,Y}$
  induce a shift in the coefficient of the nucleon scalar density $\tilde{C}_{S,Y}^{NN}$, as discussed in 
Ref.~\cite{CKOT}. We do not explicitly include this  effect  as it is not  relevant  to   our discussion.} 
\begin{equation}
\tilde{C}^{NN}_{O,Y} = \sum_{q=u,d,s}   \ G_O^{N,q}  \, C_{O,Y}^{qq}~.
\end{equation}
However, we remove the tensor operators,
because their effects  can be reabsorbed into shifts to the 
axial-vector and scalar operator coefficients. 
In fact, to leading order in a non-relativistic expansion 
$\overline{N}\sigma^{ij} N = \epsilon^{ijk}   \overline{N} \g^k \g_5 N$,
so that the spin-dependent nucleon effective Lagrangian  for $\mec$ reads 
\bea
-2\sqrt{2} G_F   \sum_{N}
  \sum_{Y}
{\Big (} 
\widetilde{C}_{A,Y}^{NN} (\overline{e} \gamma^\a P_Y \mu ) 
(\overline{N} \gamma_\a \g_5 N ) 
+h.c. {\Big ) }~~~
\label{deltaL2}
\eea
where $N\in\{ n,p\}$, $X,Y\in\{L,R\}$,  $X\neq Y$ and
\beq
\widetilde{C}_{A,Y}^{ NN} 
= \sum_q  {\Big (}  G^{N,q}_A  C_{A,Y}^{qq}
+2    G^{N,q}_T  C_{T,X}^{qq} {\Big )}~.
\label{nuclCs}
\eeq

{ Furthermore}, at finite recoil  the tensor operator induces a contribution
to the SI amplitude,  since   $\overline{u_N}(p) \sigma ^{0i} u_N(p-q)$ 
contains a term proportional to $q^i/m_N$~\cite{FHKLX,etalCirelli},  
which contracts,  in the amplitude, with the spin of the helicity-eigenstate
electron. The  net effect is tantamount to replacing the coefficient of the scalar operator with 
\beq
\widetilde{C}^{NN}_{S,Y}
\to  
\widetilde{C}^{NN}_{S,Y}
+\frac{m_\mu}{m_N}
\widetilde{C}^{NN}_{T,Y}~.
\label{recoil}
\eeq

We write 
the conversion rate 
$\Gamma = \Gamma_{SI}  + \Gamma_{SD}$,  
where $\Gamma_{SI}$ is the  $A^2$-enhanced rate occuring in
any nucleus, and $\Gamma_{SD}$ is only relevant in
nuclei with spin.  
The usual SI { branching ratio} reads~\cite{KO,KKO}
\bea
{\rm BR}_{SI} &=&   2 {\rm B}_0 \label{si}
   \bigg|  [
   \widetilde{C}^{pp}_{V,R} + \widetilde{C}^{pp}_{S,L}] \, Z \, F_p (m_\mu)
\nonumber  \\  
   & +&  [\widetilde{C}^{nn}_{V,R} + \widetilde{C}^{nn}_{S,L}]  \, [A-Z] \, F_n (m_\mu) 
\nonumber  \\
& + &    2  C_{D,L} Z e F_p (m_\mu)  
 \bigg|^2    +    \{ L \leftrightarrow R\},  \qquad
\eea
where  B$_0 =  G_F^2 m_\m^5(\a Z)^3 / ( \pi^2 \Gamma_{cap}) $, 
$ \Gamma_{cap} $ is the rate for  the  muon to transform
to a neutrino by capture
on the nucleus ($0.7054 \times 10^6$/sec in Aluminium~\cite{Suzuki:1987jf}), 
  and the  form factors $F_{p,n} (|\vec{k}|) = \int d^3x e^{-i\vec{k}\cdot \vec{x}} \rho_{p,n}(x)$  can be found in    Eq.~(30) of
Ref.~\cite{KKO}.

In  the evaluation of  $\Gamma_{SD}$ from  (\ref{deltaL2}) we treat  the muon as non-relativistic 
and the   electron as a plane wave.   Both are good approximations for
 low-$Z$  nuclei; 
  for definiteness we focus on  Aluminium ($Z=13,A=27,J=5/2$)
  the proposed target for the COMET  and Mu2e experiments. 
After approximating the muon wavefunction in the nucleus
to its  value at the origin and taking it 
outside  the  integral over the nucleus~\cite{FeinWein},
  the  nuclear part of the spin-dependent $\mu \to e$ amplitude
 corresponds to   that of  
``standard'' spin-dependent WIMP nucleus scattering.
At momentum transfer $\vec{q}$,  this is
\bea
\int d^3x e^{-i\vec{q}\cdot \vec{x}}
\langle Al |\overline{N}  (x) \g^k \g_5 N (x) | Al\rangle~.
\label{Engel}
\eea
The $\mu \to e$ amplitude is  then obtained
by multiplying by  the appropriate lepton current and coefficients   
\footnote{At  finite recoil,
 the vector or scalar operators can also
contribute to the spin-dependent amplitude~\cite{etalCirelli}.
We neglect these contributions, because we estimate
their interference with the   axial vector 
is  suppressed by ${\cal O}(m_\mu/m_N)$.}.
By analogy with WIMP scattering~\cite{EPV,BBPS,Klos:2013rwa},
we obtain:
 \bea
 {\rm BR}_{SD}  &=&
  8  {\rm B}_0
  \frac{J_{Al}+1}{J_{Al}}
\, \Big|   S^{Al}_p \widetilde{C}_{{A},L}^{ pp}  + 
 S^{Al}_n \widetilde{C}_{{A},L}^{ nn} \Big|^2    \  \frac{S_{A} (m_\mu)}{S_{A} (0)} 
\nonumber \\
&+& \{ L \leftrightarrow R \}~.
\label{ratesd}
\eea
The spin expectation values $S^{Al}_N$ are defined 
as
$S^{Al}_N = \langle  J_{Al},  J_z=J_{Al} | S_N^{z} |   J_{Al}, J_z=J_{Al}  \rangle$, 
where $S_N^z$ is the $z$ component of the  the total nucleon spin,
and the expectation value is over the nuclear ground state. 
They can be implemented in our QFT notation 
(with relativistic state normalisation for $Al$) 
by setting Eqn. (\ref{Engel}) at $|\vec{q}| = 0$  to
\bea
2    S^{Al}_N  \frac{(J_{Al})^k}{|J_{Al}|}  \times
2m_{Al}    (2\pi)^3   {\delta}^{(3)}(p_{Al,out}  - p_{Al,in}) ~. 
\nonumber
\eea
The  axial structure factor
$S_{A} (|\vec{q}|)$~\cite{EPV,Klos:2013rwa} reads
\bea
S_{A} (q) =    {a}_{L,+}^2   S_{00} (q) + 
 {a}_{L +} {a}_{L, -}   S_{01} (q) +
{a}_{L,-} ^2   S_{11} (q)  
\nonumber
\eea
where 
${a}_{L,\pm}  =  \widetilde{C}_{{A},L}^{pp} \pm \widetilde{C}_{{A},L}^{nn}$.
The  $S^{Al}_N$ and   $S_{ij}(q)$   have been
calculated in the shell model in Refs.  \cite{EngelRTO,Klos:2013rwa}.
At  $|\vec{q} | \equiv q=0$ the conversion rate is controlled by the 
spin expectation values;    we use   $S^{Al}_n=  0.030$
and  $S^{Al}_p=  0.34$~\cite{EngelRTO}.  
At finite momentum transfer $q=m_\mu$, the structure factors provide a non-trivial  correction.  
Using dominance of the proton contribution  ($S^{Al}_p >>  S^{Al}_n$) 
we find from  Ref.~\cite{EngelRTO} $S_{Al} (m_\mu) )/ S_{Al} (0)\simeq 0.29$.


{\bf  Loop effects and the RGEs}  -- 
QED and QCD loops change the magnitude
of some  operator coefficients, and 
QED  loops can transform one 
operator into another. 
 Such  Standard Model loops are
 neccessarily present, and   
  their dominant  (log-enhanced) effects   are included
 in the evolution  with
 scale  of the operator coefficients,
 as described by  the Renormalisation Group Equations (RGEs)
 of QED and QCD 
 (see \cite{Georgi} for an introduction  to
  the RG running of operators with the scale $\mu$).
If the New  Physics scale is well above
$m_W$, loops involving  the $W,Z,$ and $h$  could
also be relevant.
However,  we focus here on the RGE evolution from  
the experimental  scale $\mu_N$ 
 up to the  weak scale $m_W$.  
Since any UV model can be mapped into 
a set of operator coefficients at $\mu = m_W$,  
our calculation does not lose generality while remaining quite simple.

We consider 
the  one-loop RGEs of QED and QCD 
for $\mu \leftrightarrow e$ flavour-changing operators \cite{megmW,PSI}. 
Defining   $\lambda = \frac{ \alpha_s(m_W)}{\alpha_s(\mu_N)}$, 
their  solution can be approximated as
\bea
 C_I (\mu_N)  &\simeq & C_J(m_W)\lambda^{a_J}
 \left(
\delta_{JI} - \frac{\alpha_{e}  \widetilde{\Gamma}^e_{JI} }{4\pi} 
\log \frac{m_W}{\mu_N}  \right) ~~~
\label{oprun1l}
\eea
where $I,J$ represent the super- and subscripts which label
operator coefficients. The  
$a_I$ describe  the QCD running
 and are  only non-zero  for scalars and tensors:
for $N_f = 5$ one has  $a_I = \frac{\Gamma_{II}^s}{2\beta_0} = \{-\frac{12}{23} ,\frac{4}{23} \}$
   for $I = S,T,$.  We use this scaling  to always   give results
in  terms of coefficients at the low scale $\mu_N= 2$ GeV,
where we match quarks to nucleons.  
 ${\Gamma}^e$  is  the one-loop  
QED  anomalous dimension matrix,
rescaled ~\cite{Bellucci:1981bs,Buchalla:1989we}
 for $J,I \in T,S$ to account for the QCD
running:
\bea
\tilde{\Gamma}^e_{JI} = \Gamma^e_{JI} f_{JI}&,&f_{JI}=\frac{1}{1+a_J - a_I}
\frac{     
\lambda^{a_I - a_J} - \lambda
}{1 - \lambda}  ~.~~~
\label{fJI}
\eea
In the estimates
 presented here, we  focus on the effects
 of the off-diagonal elements of $ \widetilde{\Gamma}^e_{JI}$, 
which  mix one operator
into another, and neglect the  QED running 
of individual coefficients.

 In RG evolution down to {$\mu_N$},  
photon exchange between the external
 legs of a tensor operator can   mix it
 to  a scalar operator. 
 This contribution to the scalar coefficient  is
 \bea
   \Delta
\widetilde{C}^{NN}_{S,X }(\mu_N) \sim
\sum_q G_S^{N,q} f_{TS}
24 Q_q\frac{\alpha_e  }{\pi} 
\log \frac{m_W}{\mu_N}  C^{qq}_{T,X} (\mu_N) ~~ 
\label{TtoSsi}
\eea
where $f_{TS}$ is from Eq. (\ref{fJI}).

The tensor operator also mixes to 
the   dipole, 
when the  quark lines are closed and an
external photon is attached.  This
gives a contribution to the dipole coefficient
\bea
\left|\Delta C^{e\mu}_{D,X}(\mu_N)\right| \sim
 \frac{2 Q_q N_c m_q }{e m_\mu} \frac{\alpha_e}{\pi}
\log \frac{m_W}{\mu_N}  C^{ qq}_{T,X}(\mu_N) ~
\label{dipole}
\eea
which is suppressed  by $m_q/ m_\mu$, due
to a mass insertion on the quark line.
For tensor operators involving  $u,d$ or $s$ quark bilinears,
the mixing to the scalar operator
described in Eq. (\ref{TtoSsi})
gives a larger
contribution to SI $\mec$ than this
mixing to the dipole. So for the
remainder of this letter, we do not discuss
the contribution of   Eq. (\ref{dipole}) to
$\mec$. We will
discuss heavier quarks \footnote{The heavy quark
  scalar contribution to $\mec$\cite{CKOT} is suppressed
  $\propto 1/m_Q$, so the tensor mixing to the dipole
  could dominate.}
  in a later publication \cite{CDKS}.

Curiously, one-loop QED corrections to the axial operator generate the
vector {\cite{PSI}} \footnote{If the lepton current contained
$\g^\mu$, rather than $\g^\mu P_Y$, this would not occur.}.
If a New Physics model induces  a non-zero coefficient
$C_{A,Y}^{qq}(m_W)$, then photon exchange between the external
legs induces a  contribution to the vector coefficient
at the experimental scale:
\beq
\Delta {C}^{qq}_{V,Y}(\mu_N)
\simeq  -3 Q_q  \frac{\alpha_{e}  }{\pi} 
\log \frac{m_W}{\mu_N}
{C}^{qq}_{A,Y}(\mu_N)
\label{oprunA}
\eeq
As a result, the SI and SD processes will have
comparable sensitivities to 
axial vector operators.

{\bf Results} -- 
To interpret our  results, we first
estimate   the {\it sensitivity} of SD
and SI $\mec$   to
 the  coefficients of the  tensor and  axial operators
 of eqn (\ref{opsdefn}).
We  allow  a single 
operator coefficient  to be non-zero at $m_W$,  
and consider its  various contributions to
SD and SI $\mec$ (sometimes refered to
as setting bounds ``one-operator-at-a-time'').

Suppose first that  only the tensor coefficient ${C}_{T,L}^{uu}$
is present at $m_W$. Recall that  $C^{uu}_{T,L}(m_W)$ can
contribute to $\mec$ in three ways:  
 to the SI rate via the
finite momentum transfer effects
of eqn (\ref{recoil}),   to the SI rate  via the RG mixing
to the scalar given in eqn (\ref{TtoSsi}),
and directly to the SD rate as given in eqn (\ref{ratesd}).
It is easy to check that
the RG mixing  contribution to
$\widetilde{C}_S^{NN} (\mu_N)$ is
an order of magnitude larger   than the
finite recoil contribution. Furthermore,
the RG mixing  effect is  dominant  contribution
of  $C^{uu}_{T,L}(m_W)$ 
to $\mec$, as can 
been seen numerically
by calculating the  SD and SI contributions
to the  branching ratio:
\beq
BR (\mu{\rm Al} \to e{\rm Al}) \sim { .12} |1.54 C_{T,L}^{uu}|^2 +
 .27 |47C_{T,L}^{uu}|^2
\label{tensorsivssd}
\eeq
where  the coefficients are
  at the experimental scale, and
the second term is the $A^2$-enhanced
SI contribution.

   The RG mixing is the largest  contribution  of
 $C^{uu}_{T,L}(m_W)$ 
to $\mec$ due to three enhancements:
first, the anomalous dimension $\Gamma_{TS}^e$
is large,  and second,  the $G_S^{N,q}$ coefficients of
eqn (\ref{G}) are  an order of
magnitude larger than $G_T^{N,q}$. The combination
of these  gives 
 $ \Delta
\widetilde{C}^{NN}_{S,X }(\mu_N)
\gsim \widetilde{C}^{pp}_{T,X }(\mu_N)$,
which respectively contribute  to the SI and  SD rates.
Finally, 
the scalar coefficient benefits from
a further  $A^2$ enhancement in the SI  conversion rate.
 This shows that including the RG effects can  change
the branching ratio by  orders of magnitude.

A similar estimate for the axial operator
 ${\cal O}_{A,L}^{uu}$ 
gives 
\beq
BR (\mu{\rm Al} \to e{\rm Al}) \sim .12 |0.84 C_{A,L}^{uu}|^2 +
 .27 |.69 C_{A,L}^{uu}|^2~.
\eeq
We see that the RG mixing of
${\cal O}_{A,L}^{uu}$  into  ${\cal O}_{V,L}^{uu}$,
whose coefficient contributes  to SI $\mec$,
also gives the best sensitivity to
$ C_{A,L}^{uu}$. However, the ratio of
SI to SD  contributions is smaller
than in the tensor case, due to the
smaller anomalous dimension in eqn (\ref{oprunA}).

SI $\mec$ will
also give the best sensitivity to
tensor and axial operators involving $d$ quarks.
However,   in the case of strange quarks,
the vector
current vanishes in the nucleon, so  ${\cal O}^{ss}_{A,Y}$
only contributes to  SD $\mec$.
  The largest contribution of the strange tensor operator
is via its mixing to the scalar,
with a  sensitivity to $C_{T,X}^{ss}$
reduced by a factor $\sim G_T^{N s}/2G_T^{N u}$ with respect
to  $C_{T,X}^{uu}$. 
The strange tensor also 
mixes significantly to the dipole (see
eqn  (\ref{dipole})) which contributes to $\meg$; 
we estimate that the  sensitivity to
${C}^{ss}_{T,Y}$ of the MEG experiment
 with $BR \sim 2\times 10^{-14}$
(as expected after their upgrade),  would be comparable
to that of  COMET or Mu2e with $BR \sim$few$\times 10^{-16}$.

\begin{figure}[t]
\begin{center}
\epsfig{file=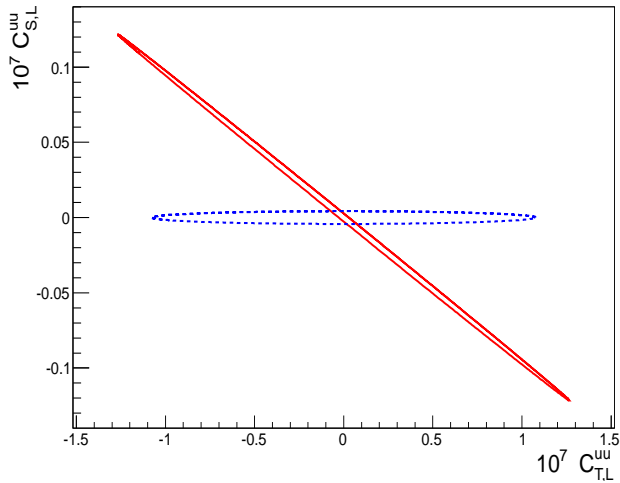,height=7cm,width=9cm}
\end{center}
\vspace{-0.4cm}
\caption{ The horizontal dotted blue (diagonal red)  
areas are the  allowed parameter space at the experimental scale
(at $m_W$), if $BR(\mec) \leq 10^{-14}$. This
plot assumes that CLFV only occurs in up-quark
operators.}
\label{fig:RGE}
\end{figure}

Let us now focus on the {\it complementarity}
of SD and SI  
contributions to the $\mec$ rate,  which
depend on different combinations of
operator coefficients. So  once
a signal is observed, measuring $\mec$
in targets with and without spin 
could  assist in differentiating among
operators or models.  
To  illustrate this complementarity,
we   restrict to scalar and tensor  operators involving $u$
quarks, whose coefficients we would like to determine. Figure \ref{fig:RGE} represents the
allowed parameter space  for  $C^{uu}_{T,L}$
and $C^{uu}_{S,L}$ evaluated at { $\mu_N$} (dotted blue) and $m_W$ (solid red).
We see that, irrespective of the operator scale, 
SD $\mec$ always gives an independent
constraint. In its absence, there
would be an unconstrained direction in parameter space, corresponding to
$C_{T,Y}^{uu}$ at the experimental scale,
or the diagonal red  { band} at $m_W$.
{ The figure also shows} that  the enhanced sensitivity
of SI conversion illustrated in eqn (\ref{tensorsivssd}) requires 
the (model-dependent) assumption that 
the model does not induce a  scalar contribution
which cancels the  mixing of the tensor into the
scalar, which would correspond to venturing
 along the red ellipse in the plot.

{\bf Prospects} -- In this letter, we followed the pragmatic 
low-energy perspective  of parametrising charged Lepton
Flavour Violating interactions with effective operators,
and considered the contribution of axial vectors and tensors
  to $\mec$. To our knowledge, this has not been 
studied previously. We found that 
the Spin-Dependent  process 
depends on  different operator coefficients  
from the Spin-Independent case, so 
comparing
$\mec$ rates in targets with and without spin 
would give additional
constraints,   and could  allow to  identify axial or tensor
operators coefficients.  
In future work\cite{CDKS},  we  plan to 
give  rates  for a complete set of
operators, 
estimate their uncertainties due to
higher order terms and neglected effects,
and    explore realistic prospects for distinguishing 
models/operators  using targets  with and without spin, 
such as different isotopes of $Ti$,  
a nucleus used  for the past  $\mec$ conversion searches.

\begin{acknowledgments}
  VC and SD thank Martin Hoferichter for discussions,  and acknowledge
  the partial support and hospitality of the Mainz Institute
  for Theoretical Physics (MITP) during the   completion
  of this work.   The work of Y.K. is supported in part
  by the Japan Society of the Promotion
of Science (JSPS) KAKENHI Grant No. 25000004.
\end{acknowledgments}


\begin{thebibliography}{222222}

\bibitem{LHC}
  V.~Khachatryan {\it et al.} [CMS Collaboration],
  ``Search for heavy Majorana neutrinos in e$^{\pm}$e$^{\pm}$+ jets and e$^{\pm}$ $\mu^{\pm}$+ jets events in proton-proton collisions at $ \sqrt{s}=8 $ TeV,''
  JHEP {\bf 1604} (2016) 169
  doi:10.1007/JHEP04(2016)169
  [arXiv:1603.02248 [hep-ex]].

  V.~Khachatryan {\it et al.} [CMS Collaboration],
  ``Search for heavy Majorana neutrinos in $\mu^\pm \mu^\pm$ jets events in proton-proton collisions at $\sqrt{s}$ = 8 TeV,''
  Phys.\ Lett.\ B {\bf 748} (2015) 144
  doi:10.1016/j.physletb.2015.06.070
  [arXiv:1501.05566 [hep-ex]].


 G.~Aad {\it et al.} [ATLAS Collaboration],
  ``Search for heavy Majorana neutrinos with the ATLAS detector in pp collisions at $ \sqrt{s}=8 $ TeV,''
  JHEP {\bf 1507} (2015) 162
  doi:10.1007/JHEP07(2015)162
  [arXiv:1506.06020 [hep-ex]].
  
\bibitem{SHiP}
  S.~Alekhin {\it et al.},
  ``A facility to Search for Hidden Particles at the CERN SPS: the SHiP physics case,''
  Rept.\ Prog.\ Phys.\  {\bf 79} (2016) no.12,  124201
  doi:10.1088/0034-4885/79/12/124201
  [arXiv:1504.04855 [hep-ph]].

  \bibitem{0nu2B}
  F.~T.~Avignone, III, S.~R.~Elliott and J.~Engel,
  ``Double Beta Decay, Majorana Neutrinos, and Neutrino Mass,''
  Rev.\ Mod.\ Phys.\  {\bf 80} (2008) 481
  doi:10.1103/RevModPhys.80.481
  [arXiv:0708.1033 [nucl-ex]].

  
\bibitem{KO}
Y.~Kuno and Y.~Okada,
  ``Muon decay and physics beyond the standard model,''
  Rev.\ Mod.\ Phys.\  {\bf 73} (2001) 151
  doi:10.1103/RevModPhys.73.151
  [hep-ph/9909265].






\bibitem{Georgi}
  H.~Georgi,
  ``Effective field theory,''
  Ann.\ Rev.\ Nucl.\ Part.\ Sci.\  {\bf 43} (1993) 209.



  


\bibitem{Babartmgteg}
  B.~Aubert {\it et al.} [ BABAR Collaboration ],
  ``Searches for Lepton Flavor Violation in the Decays 
$\tau^\pm \to e^\pm \gamma$ and $\tau^\pm \to  \mu^\pm \gamma$,''
  Phys.\ Rev.\ Lett.\  {\bf 104 } (2010)  021802.
  [arXiv:0908.2381 [hep-ex]].


\bibitem{Belletmg}
  K.~Hayasaka {\it et al.} [ Belle Collaboration ],
  ``New search for 
$\tau \to e \gamma$ and $\tau \to  \mu \gamma$,
 decays at Belle,''
  Phys.\ Lett.\  {\bf B666 } (2008)  16-22.
  [arXiv:0705.0650 [hep-ex]].

\bibitem{Belletmmm}
 K.~Hayasaka {\it et al.},
  ``Search for Lepton Flavor Violating Tau Decays into Three Leptons with 719 Million Produced Tau+Tau- Pairs,''
  Phys.\ Lett.\ B {\bf 687} (2010) 139
  doi:10.1016/j.physletb.2010.03.037
  [arXiv:1001.3221 [hep-ex]].

\bibitem{Belle-2}
  T.~Aushev {\it et al.},
  ``Physics at Super B Factory,''
  arXiv:1002.5012 [hep-ex].

  
  
  
 
\bibitem{Bertl:2006up}
  W.~H.~Bertl {\it et al.} [SINDRUM II Collaboration],
  ``A Search for muon to electron conversion in muonic gold,''
  Eur.\ Phys.\ J.\ C {\bf 47} (2006) 337.
  doi:10.1140/epjc/s2006-02582-x
C.~Dohmen {\it et al.} [SINDRUM II Collaboration],
  ``Test of lepton flavor conservation in mu ---> e conversion on titanium,''
  Phys.\ Lett.\ B {\bf 317} (1993) 631.
 
  

\bibitem{Bellgardt:1987du}
 U.~Bellgardt {\it et al.} [SINDRUM Collaboration],
 ``Search for the Decay $\mu \to 3e$,''
  Nucl.\ Phys.\ B {\bf 299} (1988) 1.
  doi:10.1016/0550-3213(88)90462-2
    
  
  
\bibitem{TheMEG:2016wtm}
  A.~M.~Baldini {\it et al.} [MEG Collaboration],
  ``Search for the lepton flavour violating decay $\mu ^+ \rightarrow \mathrm {e}^+ \gamma $ with the full dataset of the MEG experiment,''
  Eur.\ Phys.\ J.\ C {\bf 76} (2016) no.8,  434
  doi:10.1140/epjc/s10052-016-4271-x
  [arXiv:1605.05081 [hep-ex]].






\bibitem{COMET}
  Y.~Kuno [COMET Collaboration],
  ``A search for muon-to-electron conversion at J-PARC: The COMET experiment,''
  PTEP {\bf 2013} (2013) 022C01.
  doi:10.1093/ptep/pts089


\bibitem{mu2e}
 R.~M.~Carey {\it et al.} [Mu2e Collaboration],
  ``Proposal to search for $\mu^- N \to e^- N$ with a single event sensitivity below $10^{-16}$,''
  FERMILAB-PROPOSAL-0973.


\bibitem{PP}
Y.~ Kuno {\it et al.} (PRISM collaboration), ''An Experimental Search for a 
$\mu N\to e N$ Conversion at Sensitivity of the Order of
$10^{-18}$  with a Highly Intense Muon Source: PRISM'', unpublished, J-PARC LOI, 2006.

\bibitem{FeinWein}
S.~Weinberg and G.~Feinberg,
  ``Electromagnetic Transitions Between mu Meson and Electron,''
  Phys.\ Rev.\ Lett.\  {\bf 3} (1959) 111.
  doi:10.1103/PhysRevLett.3.111


\bibitem{Shanker:1979ap}
  O.~U.~Shanker,
  ``$Z$ Dependence of Coherent $\mu e$ Conversion Rate in Anomalous Neutrinoless Muon Capture,''
  Phys.\ Rev.\ D {\bf 20} (1979) 1608.
  doi:10.1103/PhysRevD.20.1608

\bibitem{czarM2}
  A.~Czarnecki, W.~J.~Marciano and K.~Melnikov,
  ``Coherent muon electron conversion in muonic atoms,''
  AIP Conf.\ Proc.\  {\bf 435} (1998) 409
  doi:10.1063/1.56214
  [hep-ph/9801218].


\bibitem{KKO}
  R.~Kitano, M.~Koike and Y.~Okada,
  ``Detailed calculation of lepton flavor violating muon electron conversion rate for various nuclei,''
  Phys.\ Rev.\ D {\bf 66} (2002) 096002
   Erratum: [Phys.\ Rev.\ D {\bf 76} (2007) 059902]
  doi:10.1103/PhysRevD.76.059902, 10.1103/PhysRevD.66.096002
  [hep-ph/0203110].

\bibitem{CKOT}
  V.~Cirigliano, R.~Kitano, Y.~Okada and P.~Tuzon,
  ``On the model discriminating power of $\mu \to  e$ conversion in nuclei,''
  Phys.\ Rev.\ D {\bf 80} (2009) 013002
  doi:10.1103/PhysRevD.80.013002
  [arXiv:0904.0957 [hep-ph]].

\bibitem{JKG}
  G.~Jungman, M.~Kamionkowski and K.~Griest,
  ``Supersymmetric dark matter,''
  Phys.\ Rept.\  {\bf 267} (1996) 195
  doi:10.1016/0370-1573(95)00058-5
  [hep-ph/9506380].

\bibitem{BBPS}
  G.~Belanger, F.~Boudjema, A.~Pukhov and A.~Semenov,
  ``Dark matter direct detection rate in a generic model with micrOMEGAs 2.2,''
  Comput.\ Phys.\ Commun.\  {\bf 180} (2009) 747
  [arXiv:0803.2360 [hep-ph]].

\bibitem{EPV}
  J.~Engel, S.~Pittel and P.~Vogel,
  ``Nuclear physics of dark matter detection,''
  Int.\ J.\ Mod.\ Phys.\ E {\bf 1} (1992) 1.
  doi:10.1142/S0218301392000023
%



\bibitem{Kosmas:1993ch}
 H.~C.~Chiang, E.~Oset, T.~S.~Kosmas, A.~Faessler and J.~D.~Vergados,
  ``Coherent and incoherent (mu-, e-) conversion in nuclei,''
  Nucl.\ Phys.\ A {\bf 559} (1993) 526.
  T.~S.~Kosmas, G.~K.~Leontaris and J.~D.~Vergados,
  ``Lepton flavor nonconservation,''
  Prog.\ Part.\ Nucl.\ Phys.\  {\bf 33} (1994) 397
  doi:10.1016/0146-6410(94)90047-7
  [hep-ph/9312217].



\bibitem{FHKLX}
  A.~L.~Fitzpatrick, W.~Haxton, E.~Katz, N.~Lubbers and Y.~Xu,
  ``The Effective Field Theory of Dark Matter Direct Detection,''
  JCAP {\bf 1302} (2013) 004
  doi:10.1088/1475-7516/2013/02/004
  [arXiv:1203.3542 [hep-ph]].

\bibitem{etalCirelli}
  M.~Cirelli, E.~Del Nobile and P.~Panci,
  ``Tools for model-independent bounds in direct dark matter searches,''
  JCAP {\bf 1310} (2013) 019
  doi:10.1088/1475-7516/2013/10/019
  [arXiv:1307.5955 [hep-ph]].
  
  
\bibitem{megmW}
  S.~Davidson,
  ``Mu to e gamma and matching at mW,''
  arXiv:1601.07166 [hep-ph].


\bibitem{PSI}
    A.~Crivellin, S.~Davidson, G.~M.~Pruna and A.~Signer,
  ``Renormalisation-group improved analysis of $\mu\to e$ processes in a systematic effective-field-theory approach,''
  arXiv:1702.03020 [hep-ph].

\bibitem{Uli}
  U.~Haisch and F.~Kahlhoefer,
  ``On the importance of loop-induced spin-independent interactions for dark matter direct detection,''
  JCAP {\bf 1304} (2013) 050
  doi:10.1088/1475-7516/2013/04/050
  [arXiv:1302.4454 [hep-ph]].

\bibitem{Crivellin:2014qxa}
  A.~Crivellin, F.~D'Eramo and M.~Procura,
  Phys.\ Rev.\ Lett.\  {\bf 112} (2014) 191304
  doi:10.1103/PhysRevLett.112.191304
  [arXiv:1402.1173 [hep-ph]].

  
\bibitem{HERMES}
  A.~Airapetian {\it et al.} [HERMES Collaboration],
  ``Precise determination of the spin structure function g(1) of the proton, deuteron and neutron,''
  Phys.\ Rev.\ D {\bf 75} (2007) 012007
  doi:10.1103/PhysRevD.75.012007
  [hep-ex/0609039].

  
\bibitem{ClatticePRL}
  T.~Bhattacharya, V.~Cirigliano, R.~Gupta, H.~W.~Lin and B.~Yoon,
  ``Neutron Electric Dipole Moment and Tensor Charges from Lattice QCD,''
  Phys.\ Rev.\ Lett.\  {\bf 115} (2015) no.21,  212002
  doi:10.1103/PhysRevLett.115.212002
  [arXiv:1506.04196 [hep-lat]].


\bibitem{Hoferichter:2015dsa}
  M.~Hoferichter, J.~Ruiz de Elvira, B.~Kubis and U.~G.~Mei§ner,
Phys.\ Rev.\ Lett.\  {\bf 115} (2015) 092301
doi:10.1103/PhysRevLett.115.092301
[arXiv:1506.04142 [hep-ph]].


\bibitem{Junnarkar:2013ac}
  P.~Junnarkar and A.~Walker-Loud,
Phys.\ Rev.\ D {\bf 87} (2013) 114510
doi:10.1103/PhysRevD.87.114510
[arXiv:1301.1114 [hep-lat]].



\bibitem{Agashe:2014kda}
  K.~A.~Olive {\it et al.} [Particle Data Group],
  Chin.\ Phys.\ C {\bf 38} (2014) 090001.
  doi:10.1088/1674-1137/38/9/090001

\bibitem{Suzuki:1987jf}
  T.~Suzuki, D.~F.~Measday and J.~P.~Roalsvig,
  ``Total Nuclear Capture Rates for Negative Muons,''
  Phys.\ Rev.\ C {\bf 35} (1987) 2212.
  doi:10.1103/PhysRevC.35.2212



\bibitem{Klos:2013rwa}
  P.~Klos, J.~Menéndez, D.~Gazit and A.~Schwenk,
  ``Large-scale nuclear structure calculations for spin-dependent WIMP scattering with chiral effective field theory currents,''
  Phys.\ Rev.\ D {\bf 88} (2013) no.8,  083516
   Erratum: [Phys.\ Rev.\ D {\bf 89} (2014) no.2,  029901]
  doi:10.1103/PhysRevD.89.029901, 10.1103/PhysRevD.88.083516
  [arXiv:1304.7684 [nucl-th]].
  
\bibitem{EngelRTO}
  J.~Engel, M.~T.~Ressell, I.~S.~Towner and W.~E.~Ormand,
  ``Response of mica to weakly interacting massive particles,''
  Phys.\ Rev.\ C {\bf 52} (1995) 2216
  doi:10.1103/PhysRevC.52.2216
  [hep-ph/9504322].


\bibitem{Bellucci:1981bs}
  S.~Bellucci, M.~Lusignoli and L.~Maiani,
  Nucl.\ Phys.\ B {\bf 189} (1981) 329.
  doi:10.1016/0550-3213(81)90384-9


\bibitem{Buchalla:1989we}
  G.~Buchalla, A.~J.~Buras and M.~K.~Harlander,
  Nucl.\ Phys.\ B {\bf 337} (1990) 313.
  doi:10.1016/0550-3213(90)90275-I


\bibitem{CDKS}
V.~Cirigliano, S.~Davidson, Y.~Kuno, A.~Saporta, work in progress.
  

  \end{thebibliography}
\end{document}